\documentclass[letterpaper]{article} 
\usepackage[preprint]{aaai2027}  
\usepackage{times}  
\usepackage{helvet}  
\usepackage{courier}  
\usepackage[hyphens]{url}  
\usepackage{graphicx} 
\usepackage{natbib}  
\usepackage{caption} 
\usepackage{algorithm}
\usepackage{algorithmic}
\usepackage{subcaption}
\usepackage{graphicx}
\usepackage{bm}
\usepackage{amsmath}
\usepackage{multirow}
\usepackage{pifont}
\usepackage[table]{xcolor}
\definecolor{tableheadercolor}{HTML}{E7D4E1} 
\definecolor{maroon}{HTML}{C00000}
\definecolor{sky_blue}{HTML}{D9E7FC}
\usepackage{arydshln}
\usepackage{booktabs}
\usepackage{enumitem}

\usepackage{amssymb}   
\usepackage{dsfont} 

\usepackage{newfloat}
\usepackage{listings}
\DeclareCaptionStyle{ruled}{labelfont=normalfont,labelsep=colon,strut=off} 
\floatstyle{ruled}
\newfloat{listing}{tb}{lst}{}
\floatname{listing}{Listing}

\title{Preference-Drift-Aware Subsequence Learning and Hierarchical Context Fusion for Long-Sequence Generative Recommendation}
\author{
    Fei Li\textsuperscript{\rm 1},
    Qingyun Gao\textsuperscript{\rm 1},
    Jianzhe Zhao\textsuperscript{\rm 1\textdagger},
    Guibing Guo\textsuperscript{\rm 1}\setcounter{footnote}{1}\thanks{Guibing Guo, Jianzhe Zhao, and Beibei Kong are corresponding authors.},
    Beibei Kong \textsuperscript{\rm 2\textdagger},
    Lei Cheng \textsuperscript{\rm 2},
    Chengxiang Zhuo \textsuperscript{\rm 2},
    Zang Li \textsuperscript{\rm 2}
}
\affiliations{
    \textsuperscript{\rm 1}Software College, Northeastern University, Shenyang, China\\
    \textsuperscript{\rm 2} Platform and Content Group, Tencent, Shenzhen,China\\

    \{neulifei,qingyungao\}@stumail.neu.edu.cn,
    \{zhaojz,guogb\}@swc.neu.edu.cn, \{echokong,raycheng,felixzhuo,gavinzli\}@tencent.com
}

\begin{document}

\maketitle

\begin{abstract}
Long-sequence generative recommendation methods autoregressively model the user’s interaction sequence to generate the next-item representation. Existing methods generally fall into two categories: efficient full-sequence modeling and target-aware context retrieval. Our experiments reveal that as the sequence length increases, the former incurs steadily growing computational cost while its accuracy gains quickly saturate and even degrade due to noise; the latter, though shortening the input sequence, is susceptible to noise that is semantically consistent yet preference-inconsistent, as well as to incomplete contexts. Both paradigms ignore the dynamic changes of user preferences and the cross-subsequence dependencies when handling historical information, thereby limiting accuracy and efficiency. To address these issues, we propose a preference-drift-aware subsequence learning and hierarchical context fusion for long-sequence generative recommendation. Specifically, we learn differentiable soft subsequence boundaries using multidimensional preference-drift information and aggregate items within each subsequence into preference-coherent representations via linear attention with soft assignment weights, thereby circumventing the expense of full-sequence attention. A cross-attention mechanism is then employed to capture dependencies between recent interactions and relevant subsequence contexts, mitigating noise in learning recent-item representations. Finally, a gated fusion mechanism adaptively combines the recent-item representation with the global subsequence context, allowing the resulting target representation to encode both recent and long-term preferences. Extensive experiments demonstrate that our method consistently outperforms existing baselines in both recommendation accuracy and computational efficiency. 
\end{abstract}

\section{Introduction}

Recommender systems are shifting from conventional multi-stage discriminative paradigms to generative recommendation systems~\cite {BLOGER}. While discriminative methods rely on query--candidate matching for next-item prediction, generative methods cast recommendation as sequence generation, autoregressively producing the next-item representation conditioned on a user’s interaction history. P5~\cite{P5} first unified diverse recommendation tasks as conditional text generation under a single language-modeling objective. To improve the semantic structure and generation efficiency of item identifiers, TIGER~\cite{TIGER} assigns each item a discrete semantic identifier (SID) via residual quantization and trains a sequence-to-sequence Transformer to generate the next item’s SID. However, this two-stage design decouples tokenization from the downstream objective, leaving the tokenizer unsupervised and potentially yielding suboptimal identifiers. DIGER~\cite{DIGER} enables recommendation-aware codebook updates via differentiable indexing and mitigates codebook collapse through annealed Gumbel noise. 
Nevertheless, these methods incur quadratic time and memory costs with respect to sequence length, limiting their ability to model complete user histories.

To address the high computational complexity of full-sequence modeling, existing long-sequence generative recommendation methods can be broadly categorized into two lines of work. (1) Efficient full-sequence modeling methods improve scalability through optimized interaction operators and attention mechanisms while preserving broad historical coverage.
For example, HSTU~\cite{HSTU} achieves efficient sequence modeling by replacing softmax attention with point-wise activation and temporal gating. RankMixer~\cite{RankMixer} improves the efficient modeling of feature interactions. However, coupling temporal signals with item semantics may cause interference. FuXi-Linear~\cite{FuXi-Linear} addresses this by decoupling temporal dynamics and positional information into dedicated channels. (2) Target-aware context retrieval methods reduce the input sequence length via interest compression or item semantic retrieval. For example, CAUSE~\cite{CAUSE} and DualGR~\cite{DualGR} compress long-term historical behaviors into a small set of interest representations based on item categories and interest disentanglement, respectively. To avoid discarding fine-grained information relevant to the current target, GLASS~\cite{GLASS} performs coarse-grained prediction on recent items via a hierarchical SID mechanism. Then it uses the resulting prediction as a dynamic key to retrieve a semantically related subset of items.

However, our assumption-validation experiments reveal that both categories of methods struggle to balance accuracy and efficiency. (1) Efficient full-sequence modeling methods incur costs that scale with sequence length but yield limited or even negative accuracy gains (Table\ref{tab:hstu_vs_fuxi}). 
(2) The performance of target-aware retrieval methods declines as sequence length increases. Although increasing the retrieval size initially improves performance, further expansion beyond a certain point degrades performance (Figure\ref{fig:GLASS_Len_Ret}). 
Overall,  their performance limitations can be attributed to the former’s homogeneous treatment of historical interactions and the latter’s reliance on retrieval based on a single similarity measure. Consequently, neither can effectively model the stage-wise evolution of user interests or the dependencies among subsequences.

To address these problems, we propose DRIFT, a preference-{D}rift-awa{R}e subsequence learning and h{I}erarchical context {F}usion me{T}hod for long-sequence generative recommendation. 
Unlike efficient full-sequence modeling with uniform aggregation and target-aware context retrieval relying on single-similarity filtering, our core idea is twofold. (1) We adaptively partition the long history into preference-consistent subsequences using preference drift information. (2) We model local- global contextual dependencies with linear complexity, thereby balancing accuracy and efficiency. Specifically, we first leverage multidimensional preference-drift information—including semantic shifts, temporal intervals, and interest distributions—to learn differentiable soft boundaries that partition a long interaction history into a few preference subsequences. Items are aggregated via linear attention weighted by item soft memberships, yielding preference-consistent subsequence representations that effectively filter historical noise irrelevant to the corresponding subsequence preference.
Secondly, we devise a hierarchical context fusion mechanism. Recent items interact with the subsequence representations through cross-attention, which incorporates context from preference-consistent subsequences while filtering out preference-inconsistent noise. A gating mechanism then adaptively fuses the local recent context with the global subsequence context, enabling the target-item representation to capture both recent and long-term preferences. Across three public datasets, DRIFT improves the performance of three backbones, achieving up to a 21.6\% gain in Recall@20 and a 97.4\% reduction in training time, demonstrating superior accuracy and efficiency in long-sequence generative recommendation.


\section{Assumption Validation}
In this section, we study how sequence length affects long-sequence generative recommendation through two experiments: (1) the impact on efficient full-sequence modeling’s accuracy and efficiency; and (2) the impact of sequence lengths and retrieval sizes on target-aware context retrieval.

\begin{table}[htbp]
\centering
\resizebox{\linewidth}{!}{   
\begin{tabular}{ll ccc ccc}
\toprule
\multirow{2}{*}{Dataset} & \multirow{2}{*}{Len.} 
& \multicolumn{3}{c}{{HSTU}} & \multicolumn{3}{c}{{Fuxi-Linear}} \\
\cmidrule(lr){3-5} \cmidrule(lr){6-8}
& & N@5 & Time (s) & \#P (M) & N@5 & Time (s) & \#P (M) \\
\midrule
\multirow{4}{*}{KuaiRec}
& 50   & 0.0676 & 1.62 & \textbf{1.04}
       & \textbf{0.0541} & \textbf{1.55} & \textbf{1.08} \\
& 100  & \textbf{0.0695} & \textbf{1.60} & 1.05
       & 0.0526 & 1.58 & 1.09 \\
& 500  & 0.0652 & 2.05 & 1.09
       & 0.0498 & 3.87 & 1.14 \\
& 1,000 & 0.0689 & 3.06 & 1.14
       & 0.0536 & 16.02 & 1.20 \\
\midrule
\multirow{4}{*}{Taobao MM}
& 50   & \textbf{0.1338} & \textbf{3.09} & \textbf{201.34}
       & \textbf{0.1340} & 17.98 & \textbf{100.71} \\
& 100  & 0.1337 & 3.99 & 201.34
       & \textbf{0.1340} & \textbf{16.52} & 100.72 \\
& 500  & \textbf{0.1338} & 10.95 & 201.38
       & 0.1339 & 36.56 & 100.77 \\
& 1,000 & \textbf{0.1338} & 21.44 & 201.43
       & 0.1339 & 609.47 & 100.84 \\
\bottomrule
\end{tabular}
}
\caption{Performance comparison between HSTU and Fuxi-Linear across different sequence lengths.}
\label{tab:hstu_vs_fuxi}
\end{table}
To assess the \textbf{accuracy and efficiency of efficient full-sequence modeling  methods}, we evaluate HSTU\cite{HSTU} and FuXi-Linear\cite{FuXi-Linear} on the KuaiRec and Taobao MM datasets across input sequence lengths of $\{50,100,500,1000\}$, while keeping other experimental settings identical to ensure a fair comparison. 
As shown in Table~\ref{tab:hstu_vs_fuxi}, increasing the sequence length yields diminishing or even negative accuracy gains while substantially increasing training  time. On KuaiRec, both methods achieve their best or near-best NDCG@5 with short sequences, whereas longer histories introduce fluctuations or degradation; on Taobao MM, NDCG@5 remains nearly unchanged across sequence lengths. In contrast, when the sequence length increases from 50 to 1,000, HSTU’s training time grows by up to 6.9×, while that of Fuxi-Linear increases by up to 33.9×. Experimental results show that, as sequence length increases, the growing number of target-irrelevant interactions degrades recommendation accuracy and computational efficiency.

\begin{figure}[htbp]
	\centering
    	\includegraphics[width=1\linewidth]{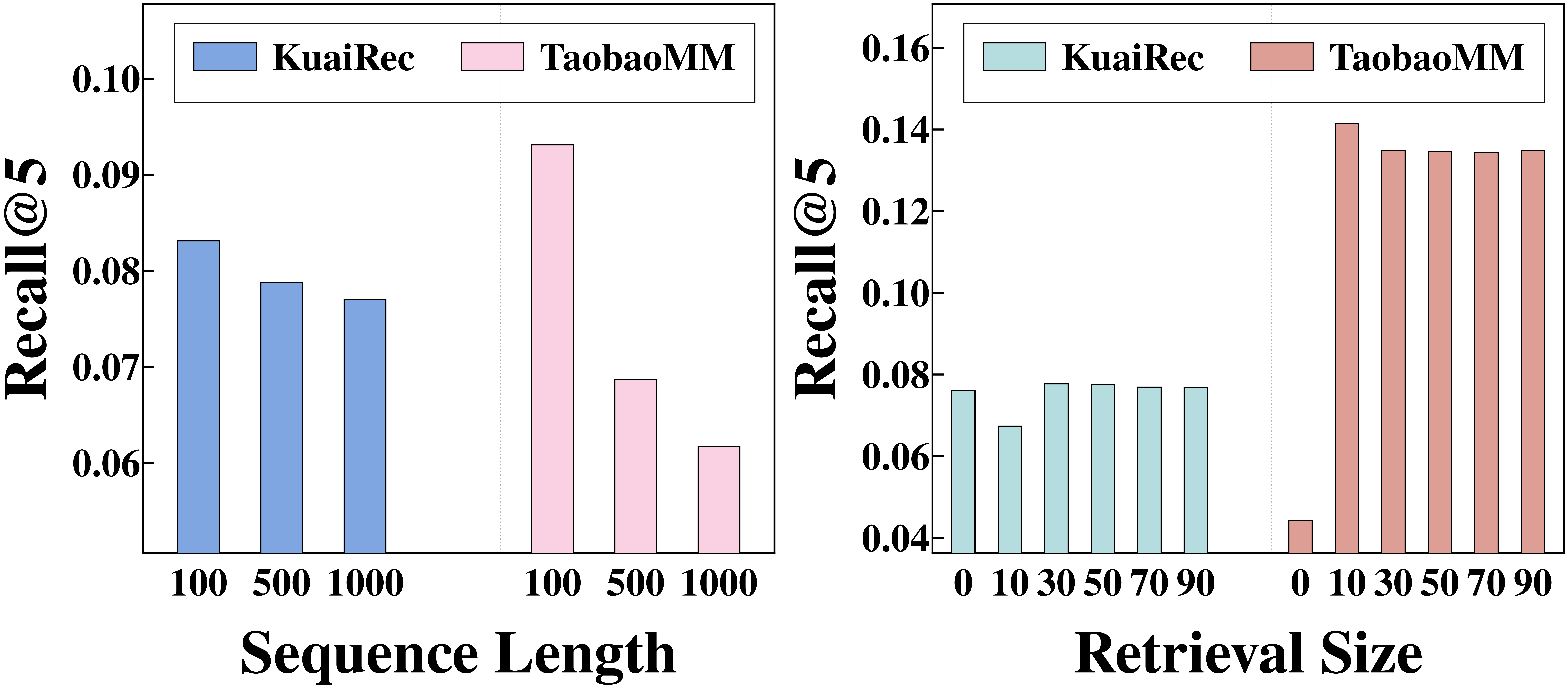}
	\caption{Performance of GLASS Across Different  Sequence Lengths and Retrieval Scales.}
	\label{fig:GLASS_Len_Ret}
\end{figure}

To assess \textbf{how sequence length and retrieval size influence target-aware context retrieval}, we evaluate GLASS~\cite{GLASS} under sequence lengths $\{100, 500, 1000\}$ and retrieval sizes $\{0, 10, 30, 50, 70, 90\}$. Appendix details experimental settings. As shown in Figure~\ref{fig:GLASS_Len_Ret}, both factors affect performance, though their impact varies across datasets. For sequence length, GLASS attains its best Recall@5 at length 100 (0.0831 on KuaiRec, 0.0931 on Taobao MM); increasing the length to 1,000 reduces Recall@5 by 7.3\% and 33.7\%, respectively. These results suggest that excessively long sequences introduce outdated or target-irrelevant interactions, thereby degrading performance.
For retrieval size, without item retrieval, Recall@5 is 0.0761 (KuaiRec) and 0.0442 (Taobao MM). On KuaiRec, performance peaks at 0.0777 with 30 items---only a marginal gain over no retrieval---while 10 items lower Recall@5 to 0.0674. On Taobao MM, retrieval yields substantial gains: Recall@5 reaches 0.1415 with 10 items, up from 0.0442 without retrieval; further increasing the retrieval size leads to a slight decline (0.1344--0.1349).
These results indicate that retrieving too few items may omit useful contextual information, whereas retrieving too many may introduce semantically similar but target-misaligned noise. Overall, relying on a single similarity metric prevents GLASS from effectively filtering such preference-inconsistent noise.

\begin{figure*}[htbp]
	\centering
    	\includegraphics[width=0.8\linewidth]{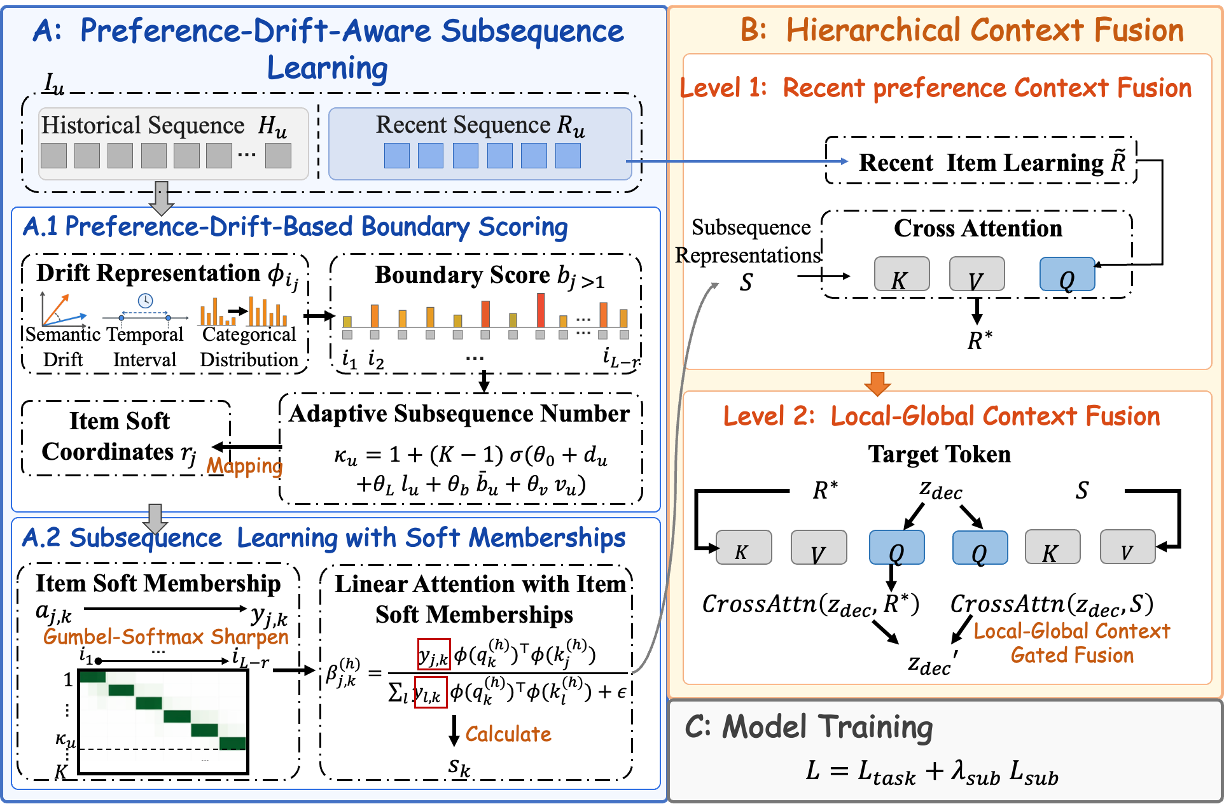}
	\caption{The overall framework diagram of DRIFT.}
	\label{fig:frame}
\end{figure*}

\section{Methodology}
We first introduce the notation and formally define generative recommendation. We then present DRIFT, which consists of a preference-drift-aware subsequence learning module and a hierarchical context fusion module, followed by the training procedure and computational complexity analysis. The overall framework is illustrated in Figure~\ref{fig:frame}.

\subsection{Preliminaries}
Let the chronologically ordered interaction sequence of user $u$ be $\mathcal{I}_u=\left\{(i_1,t_1),(i_2,t_2),\ldots,(i_L,t_L)\right\},$ where $i_j$ denotes the item involved in the $j$-th interaction and $t_j$ is the corresponding timestamp. In generative recommendation, each item $i$ is represented as a sequence of $M$ discrete tokens: $\mathbf{z}_i=\left(
z_i^{(1)},\ldots,z_i^{(M)}\right),$ where $z_i^{(m)}\in\mathcal{V}$ and $\mathcal{V}$ denotes the shared token vocabulary. Here, $M$ is the number of tokens used to represent an item. These tokens can be learned by embedding models such as BGE~\cite{BGE} and Qwen3~\cite{Qwen3}. Generative recommendation formulates next-item prediction as the autoregressive generation of the target item's token sequence:
\begin{equation}
P\!\left(i_{L+1}\mid\mathcal{I}_u\right)
=
\prod_{m=1}^{M}
P\!\left(
z_{i_{L+1}}^{(m)}
\mid
z_{i_{L+1}}^{(<m)},
\mathcal{I}_u
\right).
\end{equation}
We further divide the user interaction sequence $\mathcal{I}_u$ into a recent interaction sequence  $\mathcal{R}_u
=
\left\{
i_{L-r+1},\ldots,i_L
\right\}$, and a historical interaction sequence $\mathcal{H}_u
=
\left\{
i_1,\ldots,i_{L-r}
\right\}.$

\subsection{Preference-drift-aware Subsequence Learning}
We propose a differentiable subsequence partitioning method driven by multidimensional preference-drift information. It segments each user’s historical interaction sequence into preference-consistent subsequences, with the number of subsequences adaptively determined for each user. 

\noindent \textbf{Preference-Drift-Based Boundary Scoring.}  Given the limited ability of a single drift metric to characterize preference drift, we designed a drift characterization operator $\boldsymbol{\phi}$ that jointly models drift across three dimensions: semantic, temporal, and categorical. For user $u$’s interaction with item $i_j$ at time $t_j$, we define $\boldsymbol{\phi}_{i_j}={}\Big[\,1-\cos(\mathbf{z}_{i_j},\mathbf{z}_{i_{j-1}});\mathbf{z}_{i_j}-\mathbf{z}_{i_{j-1}};\mathrm{log}(1+\Delta t_j);\mathrm{JSD}(P_{j-w:j}\Vert P_{j:j+w})\Big].$
These four terms capture semantic dissimilarity, feature variation, interaction interval, and categorical distribution shift across the local windows around $i_j$, respectively. Subsequently, we employ a lightweight multilayer perceptron (MLP) to map the drift representation to a boundary score:
\begin{equation}
b_{j>1}=\sigma\!\big(\mathrm{MLP}_b(\boldsymbol{\phi}_{i_j})\big)\cdot \mathds{1}[j\ \text{is valid}],\qquad b_{1}=0,
\end{equation}
where $\sigma(\cdot)$ denotes the sigmoid function, and $\mathds{1}[\cdot]$ masks invalid positions. The boundary score $b_{j>1}\in[0,1]$ quantifies the likelihood that user $u$ enters a new preference stage at position $j$. A higher score indicates that the corresponding position is more likely to mark a boundary between two adjacent preference stages. Since the first position in the sequence has no preceding interaction, we set $b_{1}=0$. 

To accommodate the heterogeneity of preference-drift trajectories across users, we aggregate the position-wise boundary scores $\{b_j\}$ into three user-level statistics: the average boundary strength $\bar{b}_u$, the boundary variance $v_u$, and the normalized history length $\ell_u$. Based on these statistics, we adaptively estimate the number of subsequences:
\begin{equation} \kappa_u = 1 + (K - 1)\,\sigma\!\big(\theta_0 + d_u + \theta_L\,\ell_u + \theta_b\,\bar{b}_u + \theta_v\,v_u\big), 
\end{equation}
where $\theta_0$ and $d_u$ are the drift constants. The value $\kappa_u \in [1, K]$ grows monotonically with boundary density, drift-point concentration, and history length, enabling personalized adaptation of the subsequence number. We accumulate boundary scores into monotonically increasing soft coordinates and normalize them to the range $[1, \kappa_u]$:
\begin{equation}
r_{j}=1+(\kappa_u-1)\,\frac{\sum_{\tau\le j} b_{\tau}}{\sum_{\tau=1}^{L-r} b_{\tau}+\epsilon}.
\end{equation}

\noindent \textbf{Subsequence  Learning with Soft Memberships.} We design a soft-membership-based linear attention mechanism to learn 
preference-consistent subsequence representations. Specifically, each candidate 
subsequence $k$ is anchored to a fixed center $\mu_k = k$ and equipped with a soft 
validity gate $m_k = \sigma\big((\kappa_u - k + 0.5)/\tau_m\big)$, which smoothly 
suppresses stages beyond the user-specific horizon $\kappa_u$. We set $\tau_m = 0.3$. The membership of 
item $i_j$ in subsequence $k$ is then defined by a gated Gaussian kernel over the 
soft coordinate $r_j$:
\begin{equation} 
a_{j,k} = \frac{ m_{k}\,\exp\!\big(-(r_{j}-\mu_k)^2/\tau_a\big) }
{ \sum_{v=1}^{K} m_{v}\,\exp\!\big(-(r_{j}-\mu_v)^2/\tau_a\big) +\epsilon } . 
\end{equation}
To mitigate over-smoothing while preserving end-to-end differentiability, we further  sharpen the assignments using the Gumbel-Softmax reparameterization:
\begin{equation} 
y_{j,k} = \frac{ \exp\!\Big( \big(\log(a_{j,k}+\varepsilon)+g_{j,k}\big)/\tau_g \Big) }
{ \sum_{v=1}^{K} \exp\!\Big( \big(\log(a_{j,v}+\varepsilon)+g_{j,v}\big)/\tau_g \Big) }, 
\end{equation} 
where $g_{j,k},g_{j,v}\sim\mathrm{Gumbel}(0,1)$, and $\tau_a$ and $\tau_g$ denote the temperature 
parameters of the Gaussian kernel and the Gumbel-Softmax, respectively. Here, $\tau_a=0.5, \tau_g=1.0$.

Building upon the soft membership weights $y_{j,k}$, we define subsequence linear attention. These weights confine the representation learning of 
each subsequence to its own members, thereby avoiding cross-subsequence interference. 
Each subsequence $k$ is associated with a learnable query $\mathbf{p}_k$. Through the 
shared projections $\mathbf{q}_k = W_Q \mathbf{p}_k$, $\mathbf{k}_j = W_K \mathbf{z}_j$, 
and $\mathbf{v}_j = W_V \mathbf{z}_j$, we adopt the feature map 
$\phi(\mathbf{x}) = \mathrm{elu}(\mathbf{x}) + \mathbf{1}$ to 
compute the attention weight for head $h$:
\begin{equation}
\beta_{j,k}^{(h)}
=
\frac{
y_{j,k}\,\phi(\mathbf{q}_k^{(h)})^\top\phi(\mathbf{k}_j^{(h)})
}{
\sum_{\ell} y_{\ell,k}\,
\phi(\mathbf{q}_k^{(h)})^\top\phi(\mathbf{k}_\ell^{(h)})+\epsilon
}.
\end{equation}
The subsequence representation $\mathbf{s}_{k}$ can be calculated as 
\begin{equation}
\mathbf{s}_{k}
=
\mathrm{LN}\!\left(
\big[\,
\mathbf{o}_k^{(1)};
\,\dots;\,
\mathbf{o}_k^{(h)};
\,\dots;\,
\mathbf{o}_k^{(H)}
\,\big]
\right),
\end{equation}
where the $h$-th head output is 
$\mathbf{o}_k^{(h)}=\sum_{j=1}^{L-r} \beta_{j,k}^{(h)}\,\mathbf{v}_j^{(h)}$, and $\mathrm{LN}$ denotes the layer normalization operation.
\subsection{Hierarchical Context Fusion}
The historical subsequences $\mathbf{S}=[\mathbf{s}_1;\dots;\mathbf{s}_K]\in\mathbb{R}^{K\times d}$ encode the user's long-term preferences as they evolve, whereas the recent interaction window $\mathcal{R}_u$ captures short-term preferences. Naively concatenating the two may allow irrelevant historical subsequences to interfere with short-term preferences; conversely, focusing solely on recent interactions overlooks important long-term preferences. To address this issue, we propose a hierarchical context fusion mechanism that adaptively integrates recent and long-term preferences.

\noindent \textbf{Recent Preference Context Fusion.} We employ an autoregressive encoder (e.g., T5~\cite{T5} or HSTU~\cite{HSTU}) to learn recent item representations, i.e., 
\begin{equation}
\tilde{\mathbf{R}} = \mathbf{R}^{(i)} + f_\mathrm{A}\!\left(\mathrm{LN}_1\!\left(\mathbf{R}^{(i)}\right)\right),
\end{equation}
where $f_\mathrm{A}$ represents an attention method. In each encoder block, we use the cross-attention  to capture the correlation between recent items and subsequence representations:
\begin{equation}
\begin{aligned}
&\hat{\mathbf{R}} = \tilde{\mathbf{R}} + \mathrm{CrossAttn}\!\left(\mathrm{LN}_2\!\left(\tilde{\mathbf{R}}\right), \mathbf{K}=\mathbf{V}=\mathbf{S}\right),\\
&\mathbf{R}^{(i+1)} = \hat{\mathbf{R}} + \mathrm{FFN}\!\left(\mathrm{LN}_3\!\left(\hat{\mathbf{R}}\right)\right).
\end{aligned}
\end{equation}
 The resulting correlation weights enable each recent item to attend selectively to subsequence contexts that are consistent with its underlying preference, thereby mitigating the noise introduced by subsequences that are semantically similar but preference-inconsistent. Consequently, the hidden state $\mathbf{R}^{*}=\mathrm{LN}(\mathbf{R}^{(N)})$ encodes recent preference context.

\noindent \textbf{Local-Global Context Fusion.} We further perform adaptive context fusion at the decoder level. The decoder generates the token representations $\mathbf{z}_{\mathrm{dec}}$ of the target item in an autoregressive manner. At each decoding step, we design a gated cross-attention mechanism to fuse the recent context and global subsequence context. Specifically, this mechanism comprises two cross-attention operations. The first captures local dependencies with the recent context by attending to $\mathbf{R}^{*}$, and the second captures global dependencies across the subsequence context by attending to $\mathbf{S}$, i.e., 
\begin{equation}
\begin{aligned}
\mathbf{CA}_1 &= \mathrm{CrossAttn}\!\left( \mathbf{z}_{\mathrm{dec}}, \mathbf{R}^{*} \right),\\
\mathbf{CA}_2 &= \mathrm{CrossAttn}\!\left( \mathbf{z}_{\mathrm{dec}}, \mathbf{S} \right).
\end{aligned}
\label{eq:CA1_CA2}
\end{equation}
The resulting contextual representations are then jointly used to estimate a gate that adaptively combines recent and long-term preferences. The gate $\mathbf{g}$ dynamically controls the contribution of each context: smaller gate values favor the recent context, whereas larger values assign greater importance to the global subsequence context. Formally,
\begin{equation}
\begin{aligned}
\mathbf{g} &= \sigma\!\left( \mathbf{W}_{g} \left[ \mathbf{z}_{\mathrm{dec}}; \mathbf{CA}_2 \right] + \mathbf{b}_{g} \right),\\
\mathbf{z}_{\mathrm{dec}}' &= \mathbf{z}_{\mathrm{dec}} + \left(1-\mathbf{g}\right)\odot\mathbf{CA}_1 + \mathbf{g}\odot\mathbf{CA}_2,
\end{aligned}
\label{eq:gated_decoder_fusion}
\end{equation}
where $\odot$ denotes element-wise multiplication.

\subsection{Model Training}
Our method is applicable to target-aware context retrieval methods (e.g., GLASS) 
and efficient full-sequence modeling  methods (e.g., HSTU), unified by a loss framework:
\begin{equation}
\mathcal{L} = \mathcal{L}_{\text{task}} + \lambda_{\text{sub}}\,\mathcal{L}_{\text{sub}},\label{eq:L_main_train}
\end{equation}
where $\mathcal{L}_{\text{task}}$ is the paradigm-specific recommendation loss and $\mathcal{L}_{\text{sub}}$ is a generic subsequence structure constraint.
\begin{equation}
\mathcal{L}_{\text{sub}} = \sum_{u,k} \frac{m_{u,k}}{BK}\Big(
    \mathrm{sem}_{u,k}
    + \lambda_t \mathrm{time}_{u,k}
    + \lambda_c\,\mathrm{cat}_{u,k}
\Big),\label{eq:L_sub}
\end{equation}
where $B$ is the training batch size, and the validity gate $m_{u,k}$. $\mathrm{sem}_{u,k}$ minimizes the cosine distance to the subsequence centroid. $\mathrm{time}_{u,k}$ penalizes large gaps between adjacent items assigned to the same subsequence. $\mathrm{cat}_{u,k}$ encourages items within the subsequence to share similar categories. $\lambda_t$ and $\lambda_c$ are weighting coefficients. 

Take GLASS~\cite{GLASS} as an example, we design $\mathcal{L}_{\text{task}}$ to consist of SID generation loss, recent context recommendation loss, and subsequence recommendation loss:
\begin{equation}
\mathcal{L}_{\text{task}} = \mathcal{L}_{\text{CE}} 
+ \lambda_{\text{rec}}\,\mathcal{L}_{\text{rec}} 
+ \lambda_{\text{sub-rec}}\,\mathcal{L}_{\text{sub-rec}}.\label{eq:L_glass}
\end{equation}
SID generation loss is the autoregressive cross-entropy over the target SID tokens, given by $\mathcal{L}_{\text{CE}}=-\sum_m \log P\!\left(z^{(m)} \mid z^{(<m)},\mathcal{S}_u\right)$. $\mathcal{L}_{\text{rec}}$ uses InfoNCE to align the enhanced hidden state $\mathbf{z}_u=\operatorname{Pool}(\mathbf{R}^{*})$ with the target item embedding, while $\mathcal{L}_{\text{sub-rec}}$ applies the same objective to the subsequence representation $\mathbf{z}_{\mathrm{sub}}=\operatorname{Pool}(\mathbf{H}_{\mathrm{sub}})$.

\subsection{Time Complexity}
Let $L$ denote the sequence length, $L_h=L-r$ the length of the long-term history, $r$ the number of recent items, $K$ the number of subsequences, and $d$ the embedding dimension, where $K,r\ll L$. Our method comprises two modules: preference-drift-aware subsequence learning and hierarchical Context Fusion. The overall complexity is $\mathcal{O}\!\left(L_hKd+r^2d+rKd\right)\simeq\mathcal{O}(LKd)$. Hence, our method scales linearly with the historical sequence length $L$.

\section{Experiments}
In this section, we conduct extensive experiments to validate our methods. Due to space constraints, additional details and results are in the \textbf{Appendix}. 
\subsection{Experimental Settings}
\noindent \textbf {Description of Datasets and Baselines.} 
We conduct experiments on three public datasets: KuaiRec~\cite{KuaiRec}, ML-20M~\cite{ML-20M}, and Taobao MM~\cite{TaobaoMM}. For KuaiRec and Taobao MM, the maximum sequence length is set to 1000 with 50 recent items; for ML-20M, it is set to 200 with 10 recent items. We follow the leave-one-out protocol and use the last two interactions of each user for validation and testing. We compare our method with the following baselines: two sequential recommendation methods (GRU4Rec~\cite{GRU4Rec} and SASRec~\cite{SASRec}), two efficient full-sequence modeling  methods (HSTU~\cite{HSTU} and Fuxi-Linear~\cite{FuXi-Linear}), and three generative recommendation methods based on SIDs (TIGER~\cite{TIGER}, DIGER~\cite{DIGER}, and GLASS~\cite{GLASS}), where GLASS also belongs to the target-aware context retrieval methods.

\noindent \textbf {Implementation Details.} All methods are implemented in PyTorch and executed on NVIDIA A100 40GB. We evaluate performance using two widely adopted metrics, Recall@M and NDCG@M, with M$\in\{5, 20\}$. All methods use a hidden dimension of 96, sampled softmax with 500 negatives, and are evaluated by NDCG. SID-based generative recommendation methods use a T5-style encoder-decoder backbone with 4 encoder/decoder layers, an FFN dimension of 1024, 8 attention heads, a key/value dimension of 32, and dropout of 0.1. Moreover, for the SID-based generative recommendation methods, items are represented by three-level SIDs, with codebook sizes of [256, 256, 256] on KuaiRec and ML-20M, and [64, 128, 128] on Taobao MM. In our method, the number $K$ of subsequences is selected from $\{10,20,30\}$.  The search range of $\lambda_{\text{sub}}$ in Eq.~\eqref{eq:L_main_train} is $\{0.0, 0.02, 0.05, 0.1, 0.3\}$; those of $\lambda_t$ and $\lambda_c$ in Eq.~\eqref{eq:L_sub} are $\{0.0, 0.25, 0.5, 0.75, 1.0\}$;  and those of $\lambda_{\text{rec}}$ and $\lambda_{\text{sub-rec}}$ in Eq.~\eqref{eq:L_glass} are $\{0.1, 0.3, 0.5, 0.7, 1.0\}$. We use the AdamW~\cite{AdamW} optimizer to optimize the recommender. Models are trained for at most 100 epochs, with an early-stopping patience of 50.


\subsection{Performance Comparison}
\begin{table*}[ht]
    \resizebox{\linewidth}{!}{
        \begin{tabular}{lcccccccccccc}
            \toprule
            \multirow{2}{*}{Method} & \multicolumn{4}{c}{KuaiRec} & \multicolumn{4}{c}{ML-20M} & \multicolumn{4}{c}{Taobao MM} \\
            \cmidrule(lr){2-5} \cmidrule(lr){6-9} \cmidrule(lr){10-13}
            & N@5 & N@20 & R@5 & R@20 & N@5 & N@20 & R@5 & R@20 & N@5 & N@20 & R@5 & R@20 \\
            \midrule
            GRU4Rec         & 0.0687 & 0.0888 & 0.0892 & 0.1603 & 0.0089 & 0.0194 & 0.0155 & 0.0534 & {0.1344} & 0.1348 & 0.1350 & 0.1364 \\
            SASRec          & \underline{0.0709} & \underline{0.0924} & \underline{0.0939} & \underline{0.1703} & 0.0073 & 0.0155 & 0.0121 & 0.0419 & \underline{\textbf{0.1446}}& \underline{0.1453} & \underline{0.1455} & 0.1477 \\
            HSTU    & 0.0689 & 0.0878 & 0.0865 & 0.1540 & 0.0027 & 0.0059 & 0.0044 & 0.0158 & {0.1338} & 0.1440 & 0.1439 & 0.1443 \\

            Fuxi-Linear     & 0.0536 & 0.0670 & 0.0680 & 0.1342 & 0.0042 & 0.0110 & 0.0083 & 0.0327 & 0.1339 & 0.1341 & 0.1341 & 0.1346 \\
            TIGER           & 0.0568 & 0.0716 & 0.0735 & 0.1239 & 0.0207 & \underline{0.0396} & 0.0319 & \underline{0.1004} & 0.0012 & 0.0016 & 0.0012 & 0.0025 \\
            
            DIGER           & 0.0208 & 0.0229 & 0.0537 & 0.0631 & \underline{0.0233} & 0.0353 & \underline{0.0356} & 0.0778 & 0.0285 & 0.0517 & 0.0456 & 0.1286 \\
            GLASS           & 0.0573 & 0.0737 & 0.0777 & 0.1346 & 0.0189 & 0.0359 & 0.0303 & 0.0922 & 0.1249 & 0.1288 & 0.1365 & \underline{0.1500} \\\hdashline
            DRIFT-HSTU      & \textbf{0.0753}\textsuperscript{\dag} & \textbf{0.0993}\textsuperscript{\dag} & \textbf{0.1006}\textsuperscript{\dag} & \textbf{0.1872}\textsuperscript{\dag} & 0.0275 & 0.0480 & 0.0391 & 0.1092 & {0.1442} & 0.1445 & 0.1447 & 0.1457 \\
            DRIFT-TIGER     & 0.0618 & 0.0800 & 0.0848 & 0.1493 & \textbf{0.0349}\textsuperscript{\dag} & 0.0615 & {0.0556} & 0.1504 & 0.1136 & 0.1184 & 0.1261 & 0.1425 \\
            DRIFT-GLASS     & 0.0639 & 0.0798 & 0.0881 & 0.1532 & 0.0345 & \textbf{0.0638}\textsuperscript{\dag} & \textbf{0.0577}\textsuperscript{\dag} & \textbf{0.1608}\textsuperscript{\dag} & 0.1439 & \textbf{0.1454}\textsuperscript{\dag} & \textbf{0.1472}\textsuperscript{\dag} & \textbf{0.1524}\textsuperscript{\dag} \\
            \bottomrule
        \end{tabular}
    }
    \caption{Experimental results of 10 methods on four evaluation metrics. The best overall results are highlighted in bold, while the best results among baselines are underlined. The symbol $\dag$ indicates that our method significantly outperforms the optimal baseline at the 0.05 level based on the paired t-test.}
    \label{tab:results}
\end{table*}

\noindent \textbf{Recommendation Accuracy.}
We compare DRIFT against seven baselines across KuaiRec, ML-20M, and Taobao MM, and integrate it with HSTU, TIGER, and GLASS. The results are presented in Table~\ref{tab:results}. (1) DRIFT consistently improves accuracy across datasets and backbones. Specifically, on KuaiRec, DRIFT-HSTU performs best on all metrics, significantly improving NDCG@20 and Recall@20 over HSTU by 13.1\% and 21.6\%, respectively. On ML-20M, despite DIGER and TIGER leading the baselines, DRIFT variants top all four metrics, substantially improving every backbone with gains of up to 74.3\% over TIGER and 90.4\% over GLASS. On Taobao MM, DRIFT-GLASS ranks first on three of four metrics. 
(2)  DRIFT outperforms its backbones across HSTU, TIGER, and GLASS in all comparisons. It learns differentiable boundaries from preference-drift signals to partition long histories into preference-consistent subsequences, then integrates recent behavior and global preferences through linear attention, cross-attention, and gated fusion. This design filters irrelevant history while preserving cross-subsequence dependencies. Its larger gains on KuaiRec and ML-20M suggest greater benefits under stronger preference drift or noisier histories, while the moderate gains on Taobao MM reflect differences in data distributions and backbone strength.
\begin{table}[ht]

        \resizebox{\linewidth}{!}{
\begin{tabular}{l*{6}{c}}
\toprule
& \multicolumn{3}{c}{{KuaiRec}}
& \multicolumn{3}{c}{{Taobao MM}} \\
\cmidrule(lr){2-4} \cmidrule(lr){5-7}
{Method} & Train. & Infer. & VRAM
                & Train. & Infer. & VRAM \\
\midrule
TIGER
& 301.30 & 221.40 & 23.10
& 2743.10 & 1871.50 & 23.10 \\
+DRIFT
& \textbf{7.80} & \textbf{29.20} & \textbf{17.66}
& \textbf{34.10} & \textbf{83.40} & \textbf{16.95} \\
$\Delta$ (\%)
& -97.41 & -86.81 & -23.55
& -98.76 & -95.54 & -26.62 \\
\hdashline
GLASS
& 33.60 & 31.30 & \textbf{9.13}
& 95.50 & \textbf{60.90} & \textbf{9.13} \\
+DRIFT
& \textbf{8.70} & \textbf{31.10} & 17.75
& \textbf{34.90} & 84.80 & 17.04 \\
$\Delta$ (\%)
& -74.11 & -0.64 & +94.41
& -63.46 & +39.24 & +86.64 \\
\hdashline
HSTU
& 3.06 & 6.80 & 3.50
& 21.44 & 13.13 & 12.07 \\
+DRIFT
& \textbf{1.91} & \textbf{6.70} & \textbf{1.36}
& \textbf{12.41} & \textbf{7.41} & \textbf{7.29} \\
$\Delta$ (\%)
& -37.58 & -1.47 & -61.14
& -42.12 & -43.56 & -39.60 \\
\bottomrule
\end{tabular}
}
            \caption{Efficiency comparison across datasets. {Train.}: average training time per epoch (s); {Infer.}: inference time (s);{VRAM}: GPU memory usage (GB).}\label{tab:Efficiency_results}
	\end{table}

\noindent \textbf{Computational Efficiency.}  
By adaptively partitioning long interaction histories into a small number of preference-coherent subsequences and applying linear-complexity attention to them, DRIFT effectively filters out preference-inconsistent noise, substantially reducing training costs while accommodating much longer input sequences.  As shown in Table~\ref{tab:Efficiency_results}, DRIFT-GLASS supports sequences up to $20\times$ longer than those handled by GLASS (increasing the maximum sequence length from 50 to 1,000), while achieving $3.9\times$ and $2.7\times$ faster per-epoch training on KuaiRec and Taobao MM, respectively. DRIFT-Tiger delivers up to $80.4\times$ training speedup and $22.4\times$ inference speedup over Tiger on Taobao MM (and $38.6\times$ and $7.6\times$ on KuaiRec). Meanwhile, DRIFT-HSTU cuts per-epoch training time by 37.6\%–42.1\% and GPU memory by 39.6\%–61.1\% compared with HSTU. These results confirm that modeling compact, preference-consistent subsequences effectively mitigates the efficiency bottleneck in long-sequence recommendation.
    
\subsection{Ablation Experiments}
\begin{figure}[htbp]
    \centering
    \includegraphics[width=0.9\linewidth]{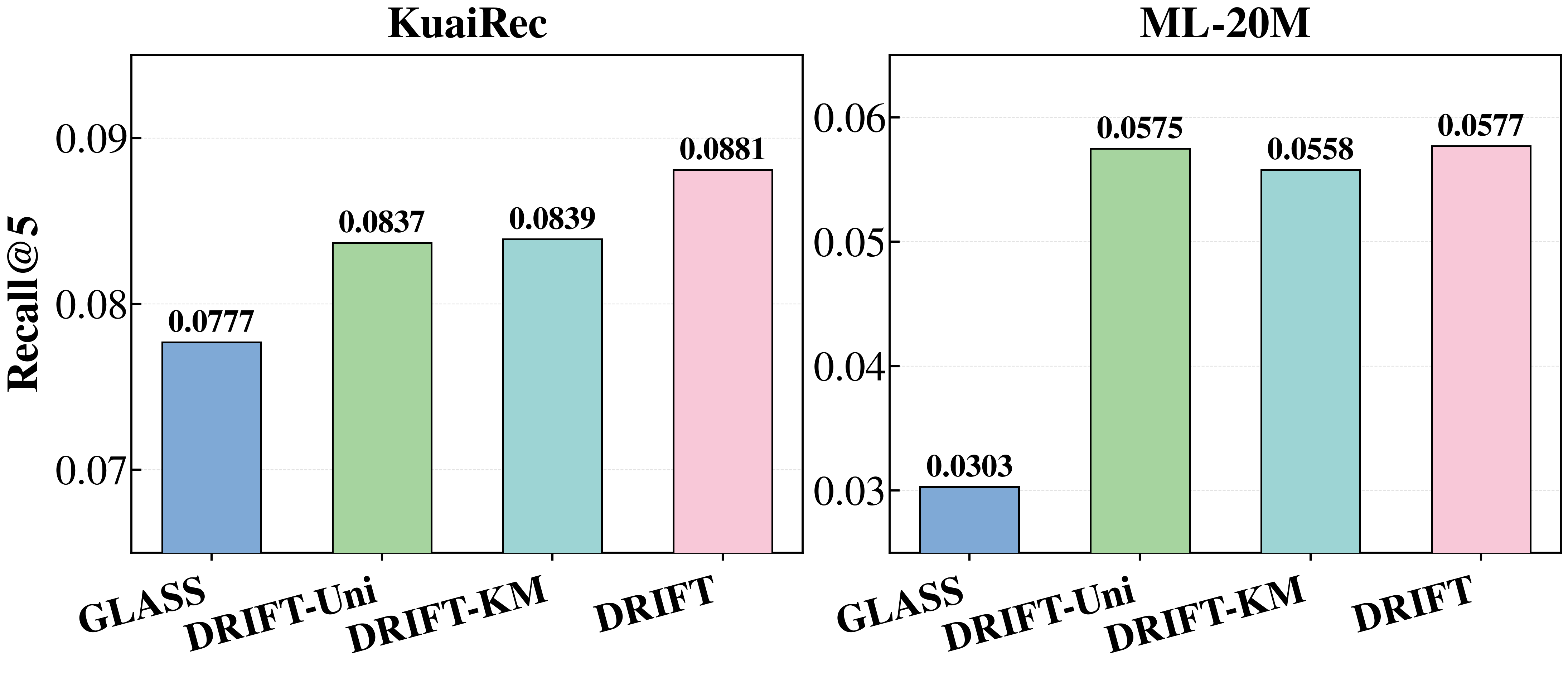}
    \caption{Performance comparison of subsequence learning.}
    \label{fig:subsequence_learning_comparison}
\end{figure}
\noindent \textbf{Effectiveness of Preference-drift-aware Subsequence Learning.} 
To validate the preference drift-aware subsequence learning module, we compare DRIFT with GLASS and two DRIFT variants: DRIFT-Uni, which uses uniform segmentation, and DRIFT-KM, which employs K-means-based segmentation. As shown in Figure~\ref{fig:subsequence_learning_comparison}, all subsequence compression-based methods outperform the retrieval-based GLASS method, indicating that compressing long sequences preserves broader historical information while mitigating context loss and noise. Among these methods, DRIFT achieves the best performance, with Recall@5 scores of 0.0881 on KuaiRec and 0.0577 on ML-20M. Unlike uniform segmentation, which overlooks the irregular evolution of user preferences, and K-means segmentation, which relies solely on representational similarity, DRIFT adaptively learns soft boundaries based on preference drift signals. This enables the construction of subsequences with greater preference consistency and effectively reduces noise in user histories, thereby demonstrating the effectiveness of preference drift-aware subsequence learning.
\begin{table}[ht]

        \resizebox{\linewidth}{!}{
        \begin{tabular}{lcccc}
\toprule
\multirow{2}{*}{Method} 
& \multicolumn{2}{c}{KuaiRec} 
& \multicolumn{2}{c}{Taobao MM} \\
\cmidrule(lr){2-3} \cmidrule(lr){4-5}
& N@5 & R@5 
& N@5 & R@5 \\
\midrule
Ours 
& \textbf{0.0639 }& \textbf{0.0881}
& \textbf{0.1439} & \textbf{0.1472} \\\hdashline

w/o Recent-CA 
& -5.32\% & -3.86\%
& -2.08\% & -0.75\% \\

Subseq-Prefix 
& -5.79\% & -5.22\%
& -99.31\% & -99.12\% \\

w/o Global-CA 
& -4.38\% & -5.45\%
& -2.02\% & -1.29\% \\

Concat-Fuse 
& -7.82\% & -5.45\%
& -99.79\% & -99.66\% \\
\bottomrule
\end{tabular}

			} 		\caption{Ablation comparison of hierarchical context fusion.}\label{tab:Ablation_hierarchical_context}
	\end{table}
\noindent \textbf{Effectiveness of Hierarchical Context Fusion.} 
As shown in Table~\ref{tab:Ablation_hierarchical_context}, DRIFT consistently outperforms all variants on both datasets, validating the effectiveness of our hierarchical context fusion mechanism. Specifically, (1) removing either \textit{w/o Recent-CA} or \textit{w/o Global-CA} consistently degrades performance, demonstrating that both local enhancement of recent preferences and long-term global context are essential; (2) directly concatenating subsequence representations, as in \textit{Subseq-Prefix} and \textit{Concat-Fuse}, leads to a catastrophic performance collapse on Taobao MM, with N@5 dropping to approximately $10^{-3}$ or lower, and also causes substantial degradation on KuaiRec. These results suggest that subsequence representations (encoding long-term preferences) and recent-item representations (capturing short-term transitions) are inherently heterogeneous. Naive concatenation introduces noise arising from semantic-space misalignment, severely disrupting the decoder representations. In contrast, our gated fusion mechanism aligns contextual information through cross-attention and mitigates conflicts between heterogeneous representations via adaptive weighting. It therefore effectively captures both recent and long-term preferences, achieving the best performance on both datasets.


\begin{figure}[htbp]
    \centering
    \includegraphics[width=0.9\linewidth]{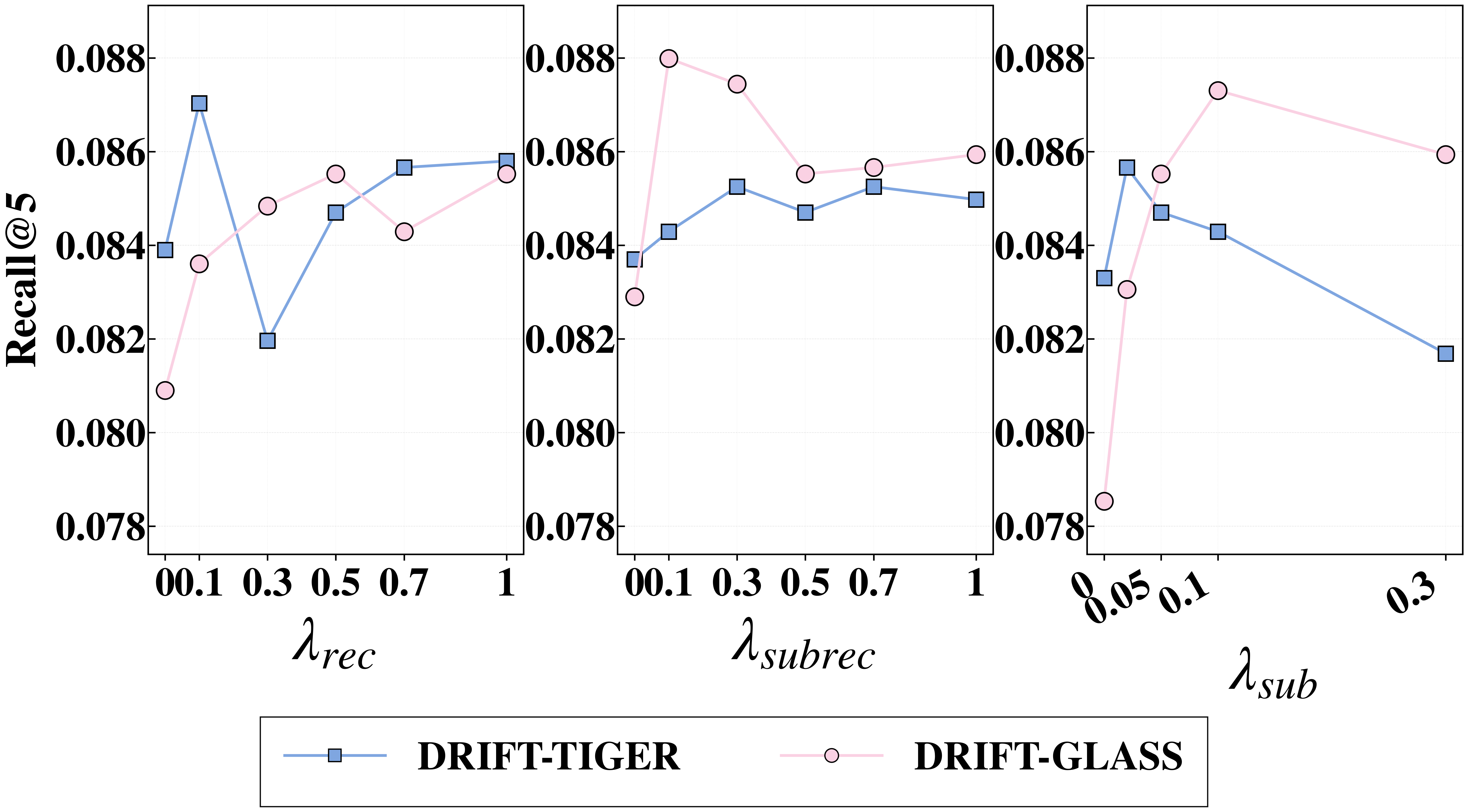}
    \caption{Performance comparison of different loss weights.}
    \label{fig:sensitivity_loss}
\end{figure}
\subsection{Sensitivity Analysis}
\noindent \textbf{Impact of Training Losses.}
We investigate the impact of $\lambda_{\text{rec}}$, $\lambda_{\text{sub-rec}}$, and $\lambda_{\text{sub}}$ on the performance of our methods using the KuaiRec dataset.
As shown in Figure~\ref{fig:sensitivity_loss}, DRIFT-TIGER achieves the best performance with $\lambda_{\text{rec}}=0.1$, $\lambda_{\text{sub-rec}}\approx0.3$, and $\lambda_{\text{sub}}\approx0.02$, while DRIFT-GLASS favors approximately $0.5$, $0.1$, and $0.1$, respectively. Excessive structural regularization degrades performance. Moreover, all three losses are beneficial. Removing $\mathcal{L}_{\text{rec}}$ causes the largest performance drop, while excluding $\mathcal{L}_{\text{sub-rec}}$ or $\mathcal{L}_{\text{sub}}$ also consistently reduces Recall@5. Overall, jointly optimizing these losses yields the best performance.

\begin{figure}[htbp]
    \centering
    \includegraphics[width=0.85\linewidth]{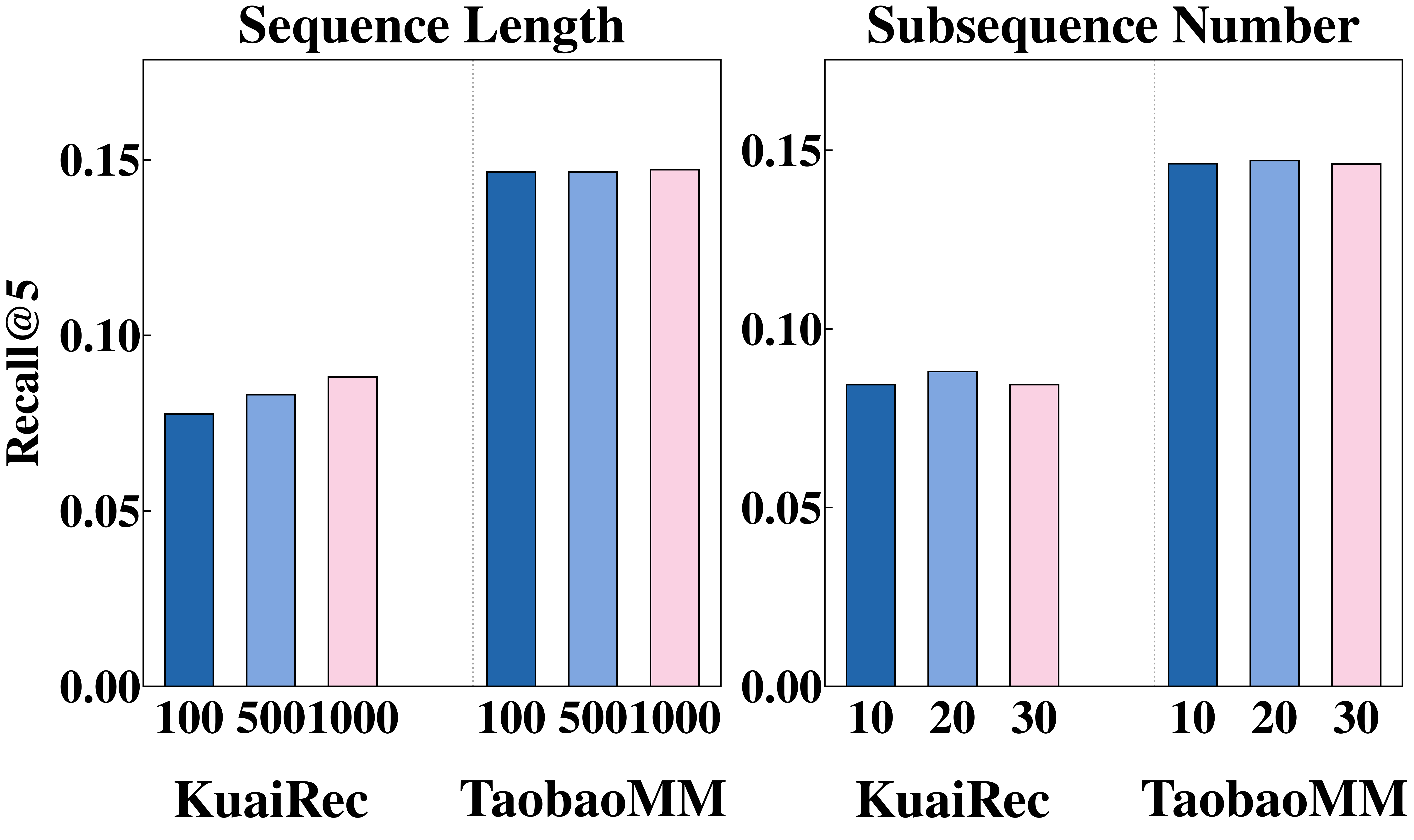}
    \caption{Performance comparison of different sequence lengths and subsequence numbers for our DRIFT-GLASS.}
    \label{fig:sensitivity_seq_len_subseq_num}
\end{figure}
\noindent \textbf{Impact of Sequence Length and Subsequence Number.}
As shown in Figure~\ref{fig:sensitivity_seq_len_subseq_num}, increasing the sequence length improves Recall@5 on KuaiRec while maintaining stable performance on Taobao MM, indicating that our method captures long-term preferences and suppresses historical noise. Both datasets peak at 20 subsequences and vary only slightly across settings. Too few subsequences may mix distinct preferences, whereas too many may fragment coherent behaviors. Overall, our method is robust to both hyperparameters.

\section{Related Work}
\noindent \textbf{Generative Recommendation.}
SID-based recommenders represent items as discrete token sequences for constrained generation. TIGER~\cite{TIGER} first quantizes item representations and then trains a Transformer to generate SIDs. To address the mismatch caused by this separate training, ETEGRec~\cite{ETEGRec} and UniSID~\cite{Unified_SID} jointly optimize tokenization and recommendation, DIGER~\cite{DIGER} further propagates recommendation gradients to the codebook, and BLOGER~\cite{BLOGER} uses bilevel optimization to balance the two objectives. This line thus evolves from content-based tokenization to recommendation-aware SID learning. Another line of research redesigns identifier generation for efficiency. SETRec~\cite{SETRec} and RPG~\cite{RPG} replace ordered SIDs with unordered tokens, enabling parallel prediction and constrained decoding. Despite advances in tokenization and generation, these methods largely rely on dense self-attention over user histories, whose quadratic cost necessitates truncation and limits long-term preference modeling.

\noindent \textbf{Long-sequence Generative Recommendation.}
Existing methods mainly fall into two categories: efficient full-sequence modeling and context reduction. The former redesigns interaction operators to preserve complete histories. HSTU~\cite{HSTU}, RankMixer~\cite{RankMixer}, OneTrans~\cite{OneTrans}, and Make It Long, Keep It Fast~\cite{LongFast} improve the scalability of long-sequence encoding through efficient architectures and training strategies. FuXi-Linear~\cite{FuXi-Linear} further decouples temporal dynamics from semantic interactions to reduce their interference. Although these methods retain broad behavioral coverage, they still process every event, including substantial irrelevant or redundant information. Context-reduction methods instead compress or retrieve historical behaviors. CAUSE~\cite{CAUSE}, GEMs~\cite{GEMs}, DualGR~\cite{DualGR}, and HiCoGen~\cite{HiCoGen} summarize histories using categories, temporal windows, disentangled interests, or hierarchical clustering. While efficient, such target-agnostic compression may discard infrequent but decisive behaviors. GLASS~\cite{GLASS} alleviates this issue by using a predicted SID to retrieve intent-relevant historical items, but inaccurate coarse predictions can exclude useful evidence. Thus, full-sequence methods preserve information at the cost of redundant computation, whereas context reduction improves efficiency at the risk of information loss.

\section{Conclusions}
We propose DRIFT to address the accuracy–efficiency trade-off and the ignore of preference drift modeling in long-sequence generative recommendation. Its core innovations include: (1) preference-drift-aware subsequence learning, which leverages multidimensional drift information to adaptively segment user histories into preference-consistent subsequences, enabling contextual aggregation and noise filtering with linear computational complexity; (2) hierarchical context fusion, which enriches recent-interaction representations through cross-attention and adaptively integrates local-global subsequence information via a gating mechanism, thereby effectively capturing both short- and long-term user preferences. Extensive experimental results demonstrate that our method outperforms existing baselines in computational efficiency and recommendation accuracy. 

\bibliography{aaai2027}


\clearpage
\appendix
\setcounter{secnumdepth}{2}
\section*{Appendix}

\section{Dataset Description} 
\begin{table}[htbp]
  \centering
  \resizebox{\linewidth}{!}{  
  \begin{tabular}{lccc}
    \toprule
    Dataset & KuaiRec & ML-20M & Taobao MM \\
    \midrule
    \#Users        & 7,176   & 138,493 & 34,535 \\
    \#Items        & 10,728  & 26,744  & 874,417 \\
    \#Interactions & 12,530,806 & 20,000,263 & 32,814,555 \\
    Sparsity & 83.72\% & 99.46\%  &  99.89\% \\
    Avg. Seq. Len. & 1,746.2 & 144.4   & 905.5 \\ 
    \bottomrule
  \end{tabular}
  }
    \caption{Statistics of the datasets.}
  \label{tab:datasets}
\end{table}
We evaluate our method on three public datasets spanning diverse recommendation domains, scales, and interaction densities: KuaiRec\cite{KuaiRec}, a short-video dataset collected from Kuaishou; ML-20M\cite{ML-20M}, a widely used MovieLens benchmark; and Taobao MM\cite{TaobaoMM}, a large-scale multimodal e-commerce dataset. Their statistics are summarized in Table~\ref {tab:datasets}. These datasets represent complementary evaluation settings. KuaiRec is relatively dense (83.72\% sparsity), with a small item catalog and exceptionally long user histories (1{,}746 interactions per user on average), making it well-suited for evaluating long-range sequential modeling. ML-20M is a medium-scale, sparse dataset (99.46\% sparsity) with moderate sequence lengths and is a standard benchmark for sequential recommendation. In contrast, Taobao MM is large-scale and highly sparse (99.89\% sparsity), with over 870K items and long behavior sequences (an average of 905 interactions). Its rich visual and textual features further enable evaluation at an industrial scale in a multimodal setting.

\textbf{Preprocessing.} For each dataset, we chronologically order user interactions and adopt the standard leave-one-out protocol: the last interaction is used for testing, the penultimate one for validation, and the remainder for training. Following common practice~\cite{SASRec,BERT4Rec}, we remove users with fewer than five interactions and truncate histories to a predefined maximum length. For Taobao MM, whose raw sequences contain up to 1{,}000 events, we additionally impose a minimum history length to ensure sufficient context.

\textbf{Semantic ID Construction.} To construct semantic identifiers (SIDs) for generative retrieval, we encode each item into a dense representation and quantize it using a residual-quantized variational autoencoder (RQ-VAE)~\cite{TIGER}. For KuaiRec and ML-20M, we encode textual content, including titles, tags, and categorical metadata, with a pretrained text embedding model BGE~\cite{BGE}. For Taobao MM, we directly use the provided pretrained multimodal item embeddings. Three-level residual quantization then maps each item to a compact hierarchical SID.
\section{Assumption Validation} 
\subsection{Experiment Setting}
To evaluate \textbf{the impact on efficient full-sequence modeling’s accuracy and efficiency}, we compare HSTU~\cite{HSTU} and FuXi-Linear~\cite{FuXi-Linear} on KuaiRec and Taobao MM with sequence lengths in $\{100, 500, 1000\}$. Both models use a single layer with a hidden dimension of $96$. HSTU~\cite{HSTU} employs one attention head, whereas FuXi-Linear~\cite{FuXi-Linear} comprises a $16$-dimensional linear temporal channel with $8$ heads and base $2$, a $16$-dimensional attention channel, and a $32$-dimensional positional channel, with a dropout rate of $0.5$. These two methods are trained for up to $100$ epochs using AdamW~\cite{AdamW}, early stopping with a patience of $50$, $500$ sampled negatives, and a softmax temperature of $0.05$. The learning rate is set to $3\times10^{-4}$ for HSTU on KuaiRec and $1\times10^{-4}$ otherwise, with a weight decay of $0.01$ on Taobao MM; FuXi-Linear uses a learning rate of $1\times10^{-4}$. Experiments use a fixed random seed of $2025$, and the best checkpoint is selected according to validation NDCG@20.

To investigate \textbf{the effects of sequence length and retrieval size on target-aware context retrieval}, we evaluate GLASS~\cite{GLASS} on KuaiRec and T ao a bo MM by varying the user-history length $L \in \{100, 500, 1000\}$ and the number of retrieved target-similar items $N \in \{0, 10, 30, 50, 70, 90\}$. Setting $N=0$ turns off the retrieval branch, reducing GLASS to a generator conditioned solely on user history. Three hierarchical codewords represent each item learned by a residual-quantized autoencoder, with codebook sizes of $[256, 256, 256]$ for KuaiRec and $[64, 128, 128]$ for Taobao MM. GLASS adopts a T5-style encoder--decoder architecture~\cite{T5} with four encoder and four decoder layers, a hidden size of $96$, a feed-forward dimension of $1024$, eight attention heads with a per-head dimension of $32$, and a dropout rate of $0.1$. The model is trained using AdamW with a learning rate of $1 \times 10^{-4}$ and a weight decay of $0.01$ for up to $100$ epochs. Early stopping is applied based on validation NDCG@20 with a patience of $50$ epochs, and predictions are generated autoregressively using beam search with a beam width of $20$. All other components and hyperparameters are fixed across experimental settings.

\begin{figure}[htbp]
    \centering
    \includegraphics[width=1\linewidth]{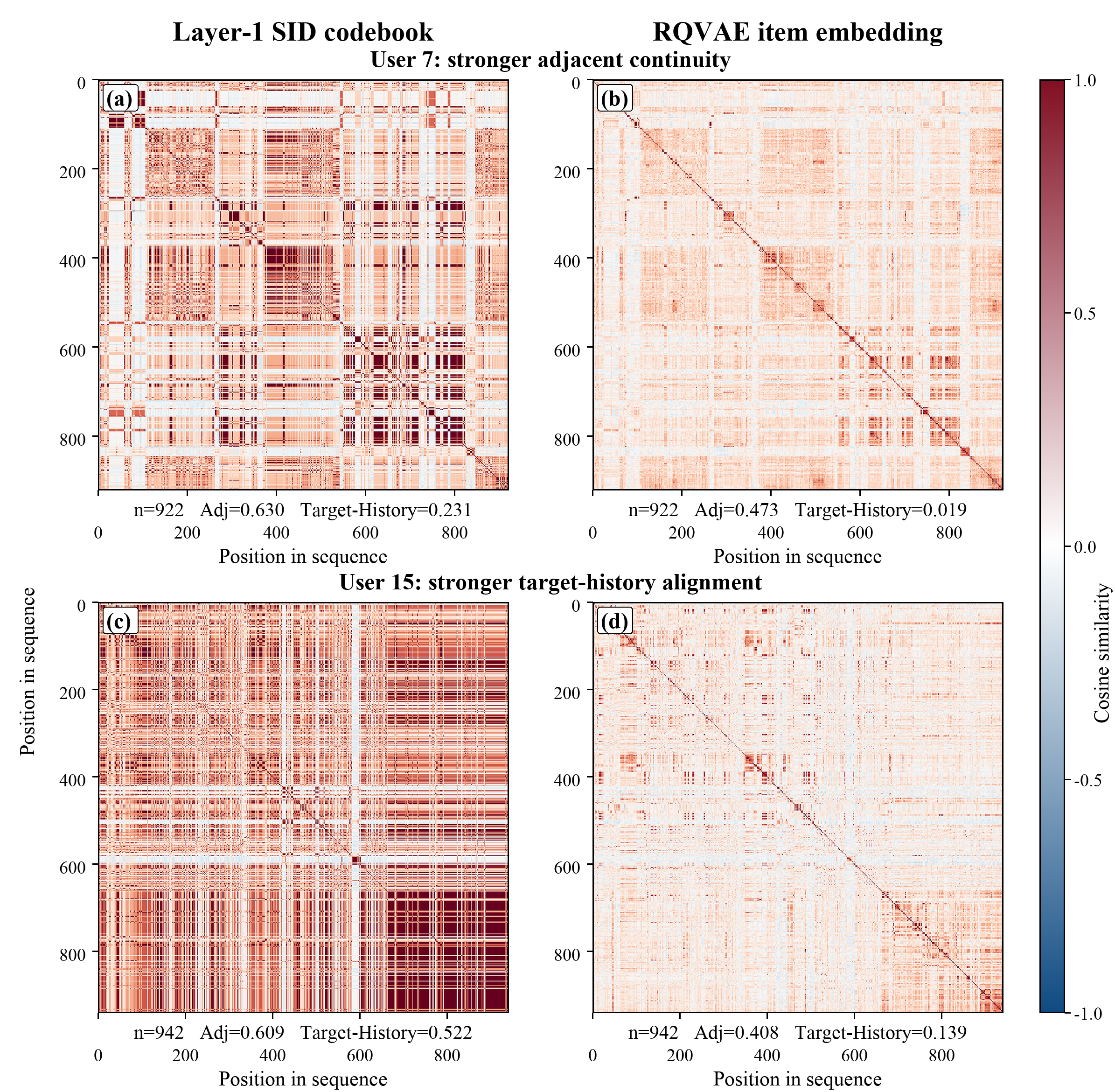}
    \caption{Heatmap of item similarity within the sequence.}
    \label{fig:taobaomm_similarity_heatmaps}
\end{figure}
\section{Interest Drift Analysis in Long Behavior Sequences}
We investigate whether long user behavior sequences comprise locally coherent, yet temporally evolving, interest subsequences in the Taobao MM dataset. As shown in Figure~\ref{fig:taobaomm_similarity_heatmaps}, both Layer-1 SID and RQVAE representations exhibit pronounced high-similarity bands along the diagonal and distinct blockwise patterns, indicating strong within-stage continuity and clear transitions across stages. For instance, User~7 demonstrates stronger continuity between adjacent interactions, whereas User~15 shows greater alignment between the target item and historical interactions. The aggregate results in Table~\ref{tab:taobaomm_local_structure} further corroborate these observations. Across 4,997 valid users, adjacent items in the original sequences achieve a mean similarity of $0.630$, exceeding those in shuffled sequences and randomly sampled item pairs by $0.392$ and $0.518$, respectively. Moreover, local coherence gradually weakens as the temporal span increases: the within-window similarity decreases from $0.630$ at $w=2$ to $0.444$ at $w=20$, while the diagonal-band gap narrows from $0.355$ at $d=1$ to $0.145$ at $d=20$. A similar pattern emerges at the subsequence level, where adjacent subsequences are consistently more similar than non-adjacent ones; for example, their similarities are $0.684$ and $0.373$, respectively, when $L=5$. In addition, the target item is more closely aligned with recent interactions than with earlier ones, yielding similarities of $0.275$ and $0.197$, respectively. Collectively, these findings suggest that long behavior sequences consist of coherent interest stages, undergo semantic transitions across stages, and exhibit a pronounced recency effect. This observation motivates the explicit modeling of dynamic interest subsequences and their contextual dependencies.


\begin{table}[htbp]
    \centering
    \resizebox{\linewidth}{!}{ 
    \begin{tabular}{lccc}
        \toprule
        \textbf{Metric} & \textbf{Observed} & \textbf{Reference} & \textbf{Diff.} \\
        \midrule
        \multicolumn{4}{l}{\textit{Adjacent-item similarity}} \\
        Adjacent vs. shuffled       & \textbf{0.630} & 0.238 & \textbf{+0.392} \\
        Adjacent vs. random pairs   & \textbf{0.630} & 0.112 & \textbf{+0.518} \\
        \midrule
        \multicolumn{4}{l}{\textit{Within-window similarity}} \\
        $w=2$                       & \textbf{0.630} & 0.238 & \textbf{+0.392} \\
        $w=5$                       & \textbf{0.561} & 0.238 & \textbf{+0.323} \\
        $w=20$                      & \textbf{0.444} & 0.238 & \textbf{+0.206} \\
        \midrule
        \multicolumn{4}{l}{\textit{Subsequence similarity}} \\
        Adjacent vs. non-adjacent, $L=5$   & \textbf{0.684} & 0.373 & \textbf{+0.311} \\
        Adjacent vs. non-adjacent, $L=10$  & \textbf{0.697} & 0.436 & \textbf{+0.261} \\
        Adjacent vs. non-adjacent, $L=20$  & \textbf{0.711} & 0.502 & \textbf{+0.209} \\
        \midrule
        \multicolumn{4}{l}{\textit{Diagonal-band structure}} \\
        Band gap, $d=1$             & \textbf{0.355} & $\approx 0$ & \textbf{+0.355} \\
        Band gap, $d=5$             & \textbf{0.245} & $\approx 0$ & \textbf{+0.245} \\
        Band gap, $d=20$            & \textbf{0.145} & $\approx 0$ & \textbf{+0.145} \\
        \midrule
        \multicolumn{4}{l}{\textit{Target--history association}} \\
        Target vs. complete history        & \textbf{0.218} & 0.114 & \textbf{+0.104} \\
        Target vs. recent/earliest five    & \textbf{0.275} & 0.197 & \textbf{+0.078} \\
        \bottomrule
    \end{tabular}
    }
        \caption{Sampled statistics of local semantic structure in Taobao MM behavior sequences.}
    \label{tab:taobaomm_local_structure}
\end{table}

\section{Model Training}
The training loss of our method consists of $\mathcal{L}_{\mathrm{task}}$ and $\mathcal{L}_{\mathrm{sub}}$:
\begin{equation}
\mathcal{L}=\mathcal{L}_{\mathrm{task}}
+\lambda_{\mathrm{sub}}\mathcal{L}_{\mathrm{sub}},
\label{eq:L_train}
\end{equation}
where $\mathcal{L}_{\text{task}}$ is the paradigm-specific recommendation loss and $\mathcal{L}_{\text{sub}}$ is a generic subsequence structure loss.

\subsection{Recommendation Loss}
Our method is applicable to target-aware context retrieval methods 
and efficient full-sequence modeling methods. Because these paradigms employ different recommendation objectives, we denote
their backbone-specific loss by $\mathcal{L}_{\mathrm{task}}$. For
DRIFT-GLASS, the recommendation loss consists of an SID generation objective
and two complementary sequence-level recommendation objectives:
\begin{equation}
\mathcal{L}_{\text{task}} = \mathcal{L}_{\text{CE}} 
+ \lambda_{\text{rec}}\,\mathcal{L}_{\text{rec}} 
+ \lambda_{\text{sub-rec}}\,\mathcal{L}_{\text{sub-rec}}.\label{app:eq:L_glass}
\end{equation}
where $\lambda_{\mathrm{rec}}$ and
$\lambda_{\mathrm{sub\text{-}rec}}$ balance the two auxiliary objectives.

\noindent\textbf{SID Generation Loss.}
The primary objective $\mathcal{L}_{\mathrm{CE}}$ is the autoregressive
cross-entropy over the target SID tokens. Let $q_{i,m}\in\{0,1\}$ indicate
whether the $m$-th target token of training instance $i$ is a non-padding
token, and let
$N_{\mathrm{tok}}=\sum_{i=1}^{B}\sum_{m=1}^{M}q_{i,m}$.
The loss is defined as
\begin{equation}
\begin{aligned}
\mathcal{L}_{\mathrm{CE}}
&= -\frac{1}{N_{\mathrm{tok}}}
\sum_{i=1}^{B}\sum_{m=1}^{M} q_{i,m} \\
&\quad{}\times \log P\!\left(
 z_i^{(m)}\mid z_i^{(<m)},\mathbf{R}^{*}_i,\mathbf{S}_i
\right).
\end{aligned}
\label{eq:L_CE}
\end{equation}
where $\mathbf{R}^{*}_i$ denotes the  hidden state with
recent context and $\mathbf{S}_i$ contains the learned historical subsequence
representations. This loss carries unit weight and provides the primary
token-level supervision for SID generation.

\noindent\textbf{Sequence-Level Recommendation Losses.}
Although $\mathcal{L}_{\mathrm{CE}}$ supervises individual SID tokens,
It does not directly enforce sequence-level alignment between a user
representation and the corresponding target item. We therefore introduce
two in-batch softmax objectives that provide complementary sequence-level
discriminative supervision.

Let
$\tilde{\mathbf{z}}_{u_i}
=\mathbf{z}_{u_i}/\|\mathbf{z}_{u_i}\|_2$
and
$\tilde{\mathbf{e}}_i
=\mathbf{e}_i/\|\mathbf{e}_i\|_2$
denote the normalized user and target-item representations, respectively.
The recent-context recommendation loss is
\begin{equation}
\mathcal{L}_{\mathrm{rec}}
=
-\frac{1}{B}\sum_{i=1}^{B}
\log
\frac{
\exp\!\left(
\tilde{\mathbf{z}}_{u_i}^{\top}
\tilde{\mathbf{e}}_i/\tau
\right)
}{
\sum_{j=1}^{B}
\exp\!\left(
\tilde{\mathbf{z}}_{u_i}^{\top}
\tilde{\mathbf{e}}_j/\tau
\right)
},
\label{eq:L_rec}
\end{equation}
where
$\mathbf{z}_{u_i}=\operatorname{Pool}(\mathbf{R}^{*}_i)$
is the pooled enhanced encoder representation, $\mathbf{e}_i$ is the
representation of the target item, and $\tau$ is the temperature.
The remaining $B-1$ target representations in the mini-batch serve as
in-batch negatives.

Similarly, let $\mathbf{z}_{\mathrm{sub},i}
=
\operatorname{Pool}
\left(
\mathbf{S}_i
\right)$
denote the representation obtained by pooling only the valid subsequence
slots. Its normalized form is
$\tilde{\mathbf{z}}_{\mathrm{sub},i}
=\mathbf{z}_{\mathrm{sub},i}/
\|\mathbf{z}_{\mathrm{sub},i}\|_2$.
The subsequence recommendation loss is
\begin{equation}
\mathcal{L}_{\mathrm{sub\text{-}rec}}
=
-\frac{1}{B}\sum_{i=1}^{B}
\log
\frac{
\exp\!\left(
\tilde{\mathbf{z}}_{\mathrm{sub},i}^{\top}
\tilde{\mathbf{e}}_i/\tau
\right)
}{
\sum_{j=1}^{B}
\exp\!\left(
\tilde{\mathbf{z}}_{\mathrm{sub},i}^{\top}
\tilde{\mathbf{e}}_j/\tau
\right)
}.
\label{eq:L_subrec}
\end{equation}
Both auxiliary objectives share the same temperature and target item
representation. In our implementation, the target representation is
retrieved from the shared SID embedding table using the first-tier SID code,
i.e.,
$\mathbf{e}_i=\operatorname{Embed}(z_i^{(1)})$.
By requiring the aggregated subsequence representation to predict the target
independently, $\mathcal{L}_{\mathrm{sub\text{-}rec}}$ discourages the
subsequence branch from degenerating into an inactive pathway.

\subsection{Subsequence Structure Loss}

The recommendation objectives supervise next-item prediction, but do not
explicitly constrain the internal coherence of the learned subsequences.
We therefore introduce $\mathcal{L}_{\mathrm{sub}}$ to encourage each
subsequence to contain semantically, temporally, and categorically
consistent interactions.

Let $a_{u,j,k}\in[0,1]$ denote the gated soft assignment of the $j$-th
historical interaction of user $u$ to subsequence $k$, where
$\sum_{k=1}^{K}a_{u,j,k}=1$. We use
\begin{equation}
m_{u,k}
=
\sigma\!\left(
\frac{\kappa_u-k+\tfrac{1}{2}}{\tau_m}
\right)
\end{equation}
as the validity gate that softly suppresses slots beyond the user-adaptive
subsequence budget $\kappa_u$. Let
$A_{u,k}=\sum_j a_{u,j,k}$ denote the soft mass assigned to subsequence $k$.
The structure constraint is
\begin{equation}
\begin{aligned}
\mathcal{L}_{\mathrm{sub}}
&= \frac{1}{B}\sum_{u=1}^{B}\sum_{k=1}^{K} m_{u,k}
\Big(\mathrm{sem}_{u,k} \\
&\qquad{}+\lambda_t\,\mathrm{time}_{u,k}
+\lambda_c\,\mathrm{cat}_{u,k}\Big).
\end{aligned}
\label{app:eq:L_sub}
\end{equation}
where $K$ is the maximum number of candidate subsequences, while
$\lambda_t$ and $\lambda_c$ control the temporal and categorical constraints.
Importantly, Eq.~\eqref{app:eq:L_sub} averages over users but sums over
subsequence slots. Because
$\sum_k m_{u,k}\approx\kappa_u$, the effective regularization budget adapts
to the number of subsequences required by each user.

\noindent\textbf{Semantic Coherence.}
We first define the assignment-weighted semantic centroid as
\begin{equation}
\bar{\mathbf{x}}_{u,k}
=
\frac{1}{A_{u,k}}
\sum_j a_{u,j,k}\mathbf{x}_{u,j},
\label{eq:sem_centroid}
\end{equation}
where $\mathbf{x}_{u,j}$ is the semantic representation of the $j$-th
historical item. The semantic constraint is
\begin{equation}
\mathrm{sem}_{u,k}
=
\frac{1}{A_{u,k}}
\sum_j a_{u,j,k}
\left(
1-
\cos\!\left(
\mathbf{x}_{u,j},
\bar{\mathbf{x}}_{u,k}
\right)
\right).
\label{eq:sem}
\end{equation}
Minimizing this term pulls co-assigned interactions toward a shared semantic
prototype, thereby discouraging a subsequence from mixing unrelated
preferences.

\noindent\textbf{Temporal Coherence.}
Let
$\Delta t_{u,j}=t_{u,j}-t_{u,j-1}$ denote the interval between two
chronologically adjacent interactions. We define
\begin{equation}
\mathrm{time}_{u,k}
=
\frac{
\sum_{j\geq 2}
a_{u,j,k}a_{u,j-1,k}
\log\!\left(1+\Delta t_{u,j}\right)
}{
\sum_{j\geq 2}
a_{u,j,k}a_{u,j-1,k}
+\epsilon
}.
\label{eq:time}
\end{equation}
The pairwise weight
$a_{u,j,k}a_{u,j-1,k}$ restricts the penalty to adjacent interactions that
are jointly assigned to the same subsequence. Meanwhile, the logarithmic
transformation limits the influence of exceptionally long inactivity
intervals. Minimizing this term, therefore, encourages preference boundaries
to align with genuine temporal discontinuities.

\noindent\textbf{Category Coherence.}
Because curated category labels are not consistently available across
datasets, we use the first-tier SID code as a learning-free proxy for the item
category. Let $\mathbf{y}^{(1)}_{u,j}$ be the one-hot first-tier SID code of
the $j$-th item and define its assignment-weighted distribution within
subsequence $k$ as
\begin{equation}
\bar{\mathbf{y}}^{(1)}_{u,k}
=
\frac{1}{A_{u,k}}
\sum_j a_{u,j,k}\mathbf{y}^{(1)}_{u,j}.
\label{eq:cat_centroid}
\end{equation}
The corresponding constraint is
\begin{equation}
\begin{aligned}
\mathrm{cat}_{u,k}
&=
\frac{1}{A_{u,k}}
\sum_j a_{u,j,k}
\mathrm{CE}\!\left(
\mathbf{y}^{(1)}_{u,j},
\bar{\mathbf{y}}^{(1)}_{u,k}
\right)\\
&=
\mathcal{H}\!\left(
\bar{\mathbf{y}}^{(1)}_{u,k}
\right).
\end{aligned}
\label{eq:cat}
\end{equation}
Thus, the weighted cross-entropy is exactly the entropy of the first-tier SID
distribution within the subsequence. Minimizing this entropy encourages each
subsequence to exhibit a concentrated coarse-semantic distribution rather
than mixing heterogeneous preference categories. All three constraints operate on the differentiable soft assignments
$a_{u,j,k}$. Consequently, their gradients propagate through the soft
partitioning process to both the boundary predictor and the adaptive
subsequence budget $\kappa_u$, enabling end-to-end learning without
subsequence boundary annotations.

\section{Experiments}
\subsection{Implementation Details.}
All models are implemented in PyTorch and trained on four NVIDIA A100 (40GB) GPUs. We use {Recall@$M$} and {NDCG@$M$} with $M\in\{5,20\}$ as the primary evaluation metrics. All methods use a hidden dimension of $d=96$.

\subsubsection{Sequence Modeling 
Baselines.}
GRU4Rec~\cite{GRU4Rec}, SASRec~\cite{SASRec}, HSTU~\cite{HSTU}, and FuXi-Linear~\cite{FuXi-Linear} are trained with sampled softmax
using 500 negative items per positive instance. HSTU and FuXi-Linear use a pointwise activation block with 1 block, 1 head, linear and attention dimensions of 16, and a dropout rate of 0.5. FuXi-Linear further decouples a
positional channel of dimension 32 and a temporal channel with 8 heads and base 2. SASRec uses 1 Transformer block, 1 head, and a dropout rate of 0.2. GRU4Rec uses a single GRU layer with hidden size 96
and dropout rate 0.2. These four methods take as input the 1,000 most recent interactions from each of the KuaiRec and Taobao-MM datasets, along with the 200 most recent interactions from the ML20M dataset. These methods are optimized with AdamW~\cite{AdamW} at
$1\times10^{-4}$, except GRU4Rec which uses Adam~\cite{Adam} at $3\times10^{-4}$. Weight decay is set to 0.01, except for GRU4Rec, where it is $1\times10^{-5}$. Models are trained for at most 100 epochs with an early stopping
patience of 50. HSTU and FuXi-Linear use a softmax temperature of 0.05.

\subsubsection{SID-based Generative  Recommendations.}
TIGER~\cite{TIGER}, DIGER~\cite{DIGER}, and GLASS~\cite{GLASS}  adopt a T5-style encoder--decoder backbone with 4 encoder layers, 4 decoder layers, 8 attention heads, key/value dimension 32, FFN dimension 1024, ReLU activation, and dropout rate 0.1. Items are encoded by three-level  SIDs, with codebook sizes $[256,256,256]$ for KuaiRec and ML-20M, and $[64,128,128]$ for Taobao MM. During inference, the next item is generated via beam search with beam size 20 over the SID vocabulary, and the top-$k$ results are
evaluated for $k\in\{5,20\}$. The model context length is fixed to 50 item tokens. The historical retrieval range $H$ is set to 1000 on KuaiRec and Taobao MM, and 200 on ML-20M; the recent window size $r$ is 50 on KuaiRec/Taobao MM and 10 on ML-20M. All models are optimized with
AdamW~\cite{AdamW} using a learning rate of $1\times10^{-4}$, trained for up to 100 epochs with early stopping patience 50.

\subsubsection{Hyperparameters of Our Method.}
The number of subsequences $K$ is selected from
$\{10,20,30\}$. The search space of $\lambda_{\text{sub}}$ in
Eq.~\eqref{eq:L_train} is $\{0.0,0.02,0.05,0.1,0.3\}$; those of
$\lambda_t$ and $\lambda_c$ in Eq.~\eqref{app:eq:L_sub} are
$\{0.0,0.25,0.5,0.75,1.0\}$; and those of $\lambda_{\text{rec}}$ and
$\lambda_{\text{sub-rec}}$ in Eq.~\eqref{app:eq:L_glass} are
$\{0.1,0.3,0.5,0.7,1.0\}$. The in-batch softmax recommendation losses
$L_{\text{rec}}$ and $L_{\text{sub-rec}}$ share a temperature
$\tau_{\text{rec}}=0.07$. The subsequence
structure loss $L_{\text{sub}}$ is normalized only by the batch size $B$.
The default search initialization is
$\lambda_{\text{sub}}=0.05$, $\lambda_t=\lambda_c=0.5$, and
$\lambda_{\text{rec}}=\lambda_{\text{sub-rec}}=0.5$. Our DRIFT is optimized
with AdamW~\cite{AdamW} and trained for up to 100 epochs with early
stopping patience 50.

\begin{figure}[htbp]
    \centering
    \includegraphics[width=1\linewidth]{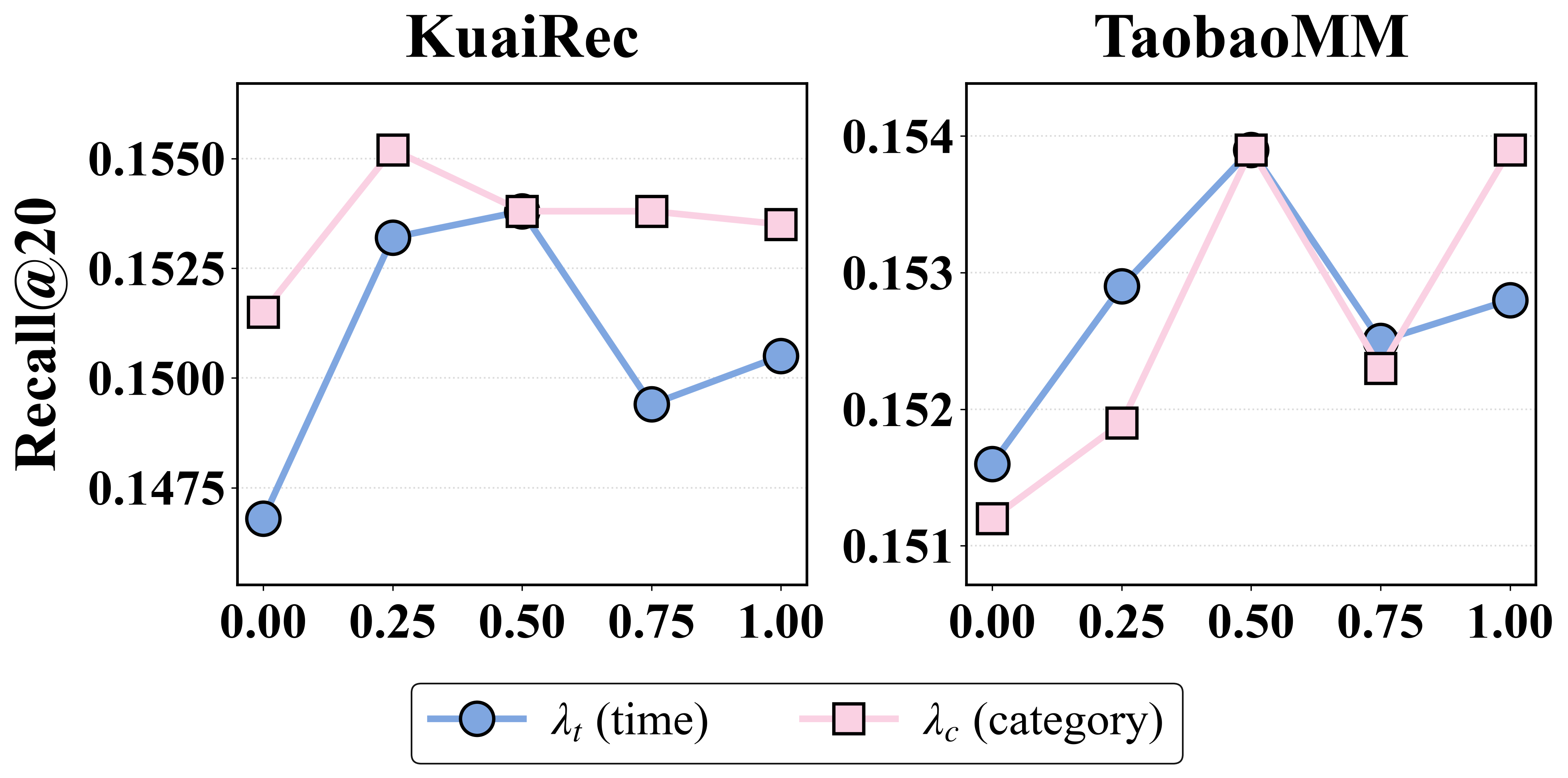}
    \caption{Performance comparison of different sequence lengths and subsequence numbers for our DRIFT-GLASS.}
    \label{fig:sensitivity_lamt_lamc}
\end{figure}
\subsection{Impact of Subsequence Constraint Weights ($\lambda_t$, $\lambda_c$)}
We analyze the sensitivity of DRIFT-GLASS to the internal weights
$\lambda_t$ and $\lambda_c$ in $\mathcal{L}_{\text{sub}}$
(Eq.~\eqref{app:eq:L_sub}). Specifically, we sweep each weight over
$\{0, 0.25, 0.5, 0.75, 1.0\}$ while fixing the other to the default value of
$0.5$. As shown in Figure~\ref{fig:sensitivity_lamt_lamc}, DRIFT-GLASS remains
highly robust to both hyperparameters. On KuaiRec, Recall@20 varies only within
$[0.147, 0.154]$ when tuning $\lambda_t$ and within $[0.152, 0.155]$ when tuning
$\lambda_c$. On TaobaoMM, the fluctuation is even smaller, with an absolute
variation no greater than $0.003$. The default setting
$\lambda_t=\lambda_c=0.5$ consistently achieves near-optimal performance across
datasets. Moreover, completely disabling either constraint term
(i.e., setting the corresponding weight to $0$) leads to 
performance degradation. These results indicate that temporal coherence and
category consistency provide complementary but non-critical regularization
effects. The observed insensitivity further suggests that the semantic coherence term $\mathrm{sem}$, whose implicit weight is $1.0$, plays the
dominant role in $\mathcal{L}_{\text{sub}}$.

\end{document}